# Dynamic Characteristics of Proton-Proton Collisions

F. H. Sawy, M. T. Ghoneim and M. T. Hussein
Physics Department, Faculty of Science,
Cairo University, 12613, Giza – Egypt
fatma.helal.sawy@cern.ch , ghoneim@sci.cu.edu.eg and
tarek@Sci.cu.edu.eg

## Abstract

The fact that quark-quark, quark-gluon and gluon-gluon interactions are different sources of particle production in proton-proton collision at high energy is investigated in this work. The variation of the multiplicity and pseudo-rapidity distributions of the created particles with the interaction energy is studied. The limits at which Feynman and KNO scaling are violated and their relations with the mechanism of the production sources are considered in view of some theoretical aspects.

## I- Introduction

The mean multiplicity of particles in the final state within different rapidity intervals has been always a good tool to probe the mechanism(s) of hadron production in high energy nuclear reactions. The rapidity of the produced particles has been also a guide for many researchers to describe the strong interaction dynamics, partially by Perturbative Quantum Chromo-Dynamics theory (PQCD) and partially by Non-Perturbative models. Parameterization of the physical quantities helped in interpretation of the hadronization of the partonic final states and the diffractive dissociation processes [1, 2].

As the energy goes from low to high values, experimental data show violation of some regimes concerned with rapidity distribution (Feynman scaling), the Koba, Nielsen, and Olsen, (KNO) scaling of multiplicity.

In the following section, we shall present a study of the variation of these parameters for the proton-proton interactions to identify the sources of hadron production at different energy. One may divide the interaction energy into three domains; namely, the Intersecting Storage Rings (ISR) up to $\sqrt{s}$ = 63 GeV, the Super Proton Synchrotron (SPS) up to $\sqrt{s}$ = 900 GeV and the Large Hadron Collider (LHC) up to $\sqrt{s}$ = 7 TeV, where s is the square of the center of mass energy of the interacting protons.

## II- Data and analysis:

In the limit, where the particle is travelling pretty close to the speed of light, it is reasonable to replace its rapidity with the pseudo-rapidity:

$$\eta = -\ln[\tan(\frac{\theta}{2})] \qquad (1)$$

, θ is angle of emission of the produced particle relative to the incident beam. The pseudo-rapidity is easy to measure because it does not require the knowledge of the particle momentum.

The pseudo-rapidity distribution of the created particles over a wide range of interaction energy [3-5] is presented in figure 1, where it seems to exhibits a couple of characters; namely:

# Constant value (plateau) around the central region (η=0) within the ISR range. The height of this plateau rises, a result of the increase in multiplicity, and its width gets wider, a result of the narrowing of the angle of the cone within which these particles are emitted, as the interaction energy rises. It is to be noted here that the rise and the widening of the plateau take place without a change in its shape. This experimental fact might recommend the presence of a single dominant source of particle production, named after as, the "soft mechanism". This source originates from fragmentation of the valance quarks, where the created particles are emitted with equal probability within a relative small angle of emission [1,6,7]. In this mechanism, the final-state particles are

created and emitted uncorrelated and it is probably the dominant mechanism in particle creation up to several tens of GeVs.

The one dominant source of creation, in this energy range, is supported by some experimental observations [2,7] as well as some theoretical concepts [8-10] in that energy domain; of these:

-The multiplicity, n, of the created particles shows a Poisson distribution [1]:

$$P(n) = \frac{<n>^n}{n!} e^{-<n>} \qquad (2)$$

-The dependence of the average multiplicity on the center of mass energy, s, shows a linear logarithmic relationship, as shown by the linear portion of figure 3[7].

-The multiplicity distributions of the produced particles follow the KNO scaling, figure 4. This is a well-established empirical law that relates the parameter, n/<n> to the quantity P(n)<n>. This scaling behavior assumes a single particle creation mechanism [11-13].

-In 1969, Feynman pointed out that the multiplicity of the produced particles becomes equally distributed in full rapidity space, so that the number of produced particles per rapidity interval, $(dn/dy)$ is independent of the collision energy:

$$\frac{dn}{dy} = \text{constant} \qquad (3)$$

Feynman considered that the maximum reachable rapidity, $y_{max}$, in nucleon-nucleon collisions increases with ln (√s):

$$y_{max} = \ln\left(\frac{\sqrt{s}}{m_n}\right) \qquad (4)$$

, where, $m_n$ is the nucleon-mass [1,14].

\# In figure 1 one would notice also that, as energy goes higher, two humps with relative large values of the pseudo-rapidity, start to show up surrounding a valley in between, around the central part. The width of the valley continues getting wider and each hump goes higher and narrower as the energy increases.

The two humps are natural results of the higher rapidity density with the rise of the energy and the narrower cone of emission of the created particles with respect to the collision axis of the two protons. Higher humps and their narrower width are two characteristics of the laser-like behavior of the created particles that become more coherent and more unified in direction, being emitted from one source. This source is still the fragmentation of the valence quarks. The wide central region would be attributed mainly to the gluon fragmentation mechanism in which no pronounced peaks are expected to show up. This is because the created particles that result from this mechanism are emitted with increased multiplicity but at relative broad angles [15-16].

This trend is confirmed by the data and calculations in figure 2. The data (given as blob points) in this figure are taken at energies of 0.9, 2.36 and 7 TeV, respectively, while solid lines are calculations according to Relativistic Diffusion Model (RDM) with three sources for particle production. The upper most solid line is a calculation according to the model predicted at 14TeV interaction energy (15). The RDM assumes that the rapidity distribution of the produced particles stems from an incoherent superposition of the beam – like fragmentation components at larger rapidity originating mostly from the valence quark-gluon interactions, and a component centered at mid-rapidity that is essentially due to gluon-gluon collisions. All of these distributions are broadened in rapidity space as a consequence of diffusion – like processes.

The increased dominance in particle creation via this mechanism might be clear from figure 3 in which, the gap in average multiplicity between the linear part and the quadratic one gets larger with the increase of the interaction energy.

Thus, once the interaction energy goes beyond the ISR limit, other sources have to be added to the soft mechanism stated above, These mechanisms share in particle creation through the sea quarks and hard gluon fragmentations, namely the semi-hard and the hard ones[1,17-19], where their contribution increases with the rise of energy. Experimentally, semi-hard events are responsible for a "mini-jet" production [23]. A "mini-jet" is defined as a group of particles having a total transverse momentum larger than 5 GeV/c [17,19]. Hard events originate from high energy quark-quark collisions which can be mathematically described by PQCD. Hard quark interactions develop via short-distance over a very short time scale and the subsequent fragmentation produces a cone of hadronic final states that originate from the same quarks. This cone of hadrons is called a "jet", representing an independent fireball for hadron creation, and the properties of the jet depend only on the initial quark. The growth of multiplicity at higher energy can be understood by assuming that; as the interaction energy goes higher; gluon jets grow with higher multiplicity than the quark jets and both compete with each other to, eventually, produce particles through their multi-fragmentations. QCD predicts that gluon initiates jets to have higher average particle multiplicity compared to quark initiated ones [16].

The first signs of the presence of such additional mechanisms showed up in the experimental data in the violations of the KNO and Feynman scaling [1,21-23]. Violation of Feynman scaling might be clear in figure 5 that shows the pseudo-rapidity distributions, around the central region, over the whole available data of interaction energy. In this figure, the particle density per unit rapidity interval is no longer constant as one goes beyond the ISR limits of energy.

One more phenomenological fact that supports the existence of other sources than valence quarks; is the description of the multiplicity distributions with single negative binomial distribution (NBD):

$$P(n;<n>;k) = \binom{n+k-1}{n}\left(\frac{<n>/k}{1+<n>/k}\right)^n \frac{1}{(1+<n>/k)^k} \qquad (6)$$

, where <n> is the mean multiplicity and k is a parameter that determines the width of the distribution, respectively. NBD is a statistical character of random events.

For (1/k) → 0, the NBD reduces to the Poisson distribution, and for k=1 it is a geometric distribution. (1/k) was found to increases linearly with ln(s) whereas KNO scaling corresponds to energy-independence. As energy increases above 900 GeV, single NBD fails to describe the multiplicity distribution and was replaced by double NBD.

Finally, one may figure out three sources of particle creation, indicated by both the experimental results and theoretical models; the soft, the semi-hard and the hard mechanisms. These mechanisms show different contributions as the interaction energy rise from the ISR up to the LHC ranges.

# Conclusion:

The experimental data that include the multiplicity and rapidity of the created particles resulting from a wide range of interaction energy of proton–proton collisions have revealed more than one source of particle creation, summarized as:

\# Soft mechanism, the source of which is the interaction of the valence quarks, where the final-state particles are created and emitted uncorrelated, and it seems to exist all over the considered range of energy. However, it is probably the only mechanism in particle creation up to several tens of GeVs.

\# Semi-hard mechanism, resulting from the sea quarks and the quark-gluon interactions, which starts to show up as energy passes through several hundreds of GeVs region. The created particles have the ability to emit additional creation by decay and cascade production.

\# Hard mechanism which starts to manifest itself on approaching the borders of the TeV region, where it takes a leading role in creation in addition to the above two components, as a result of the growth of the gluon-gluon interactions, besides the high energy quark interactions. The creation of particles via this mechanism seems to be characterized by a high multiplicity and a broad angular distribution.

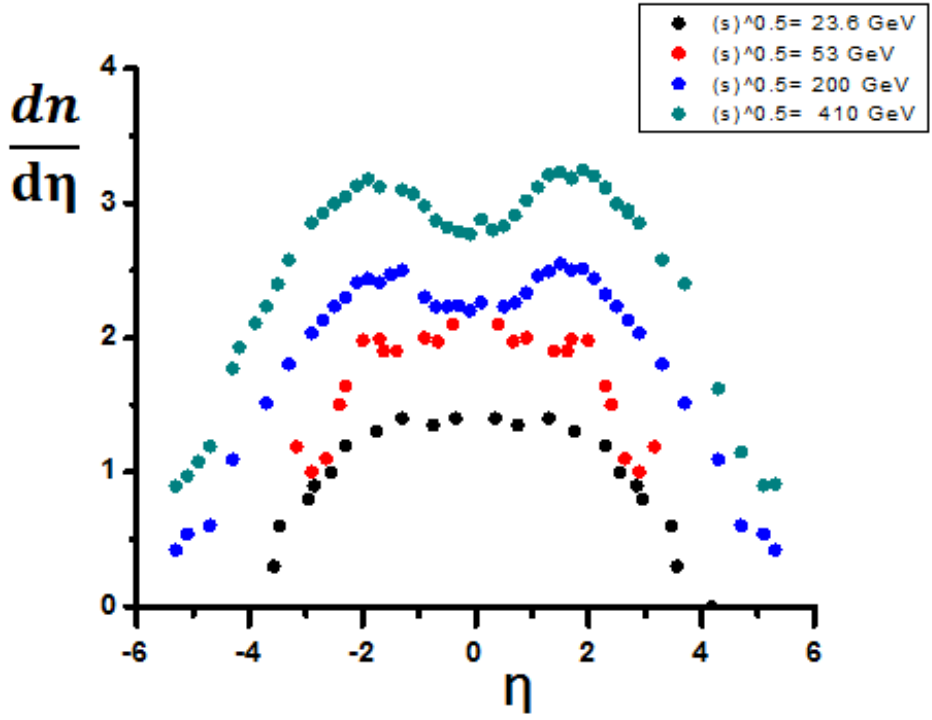

**Figure (1)**

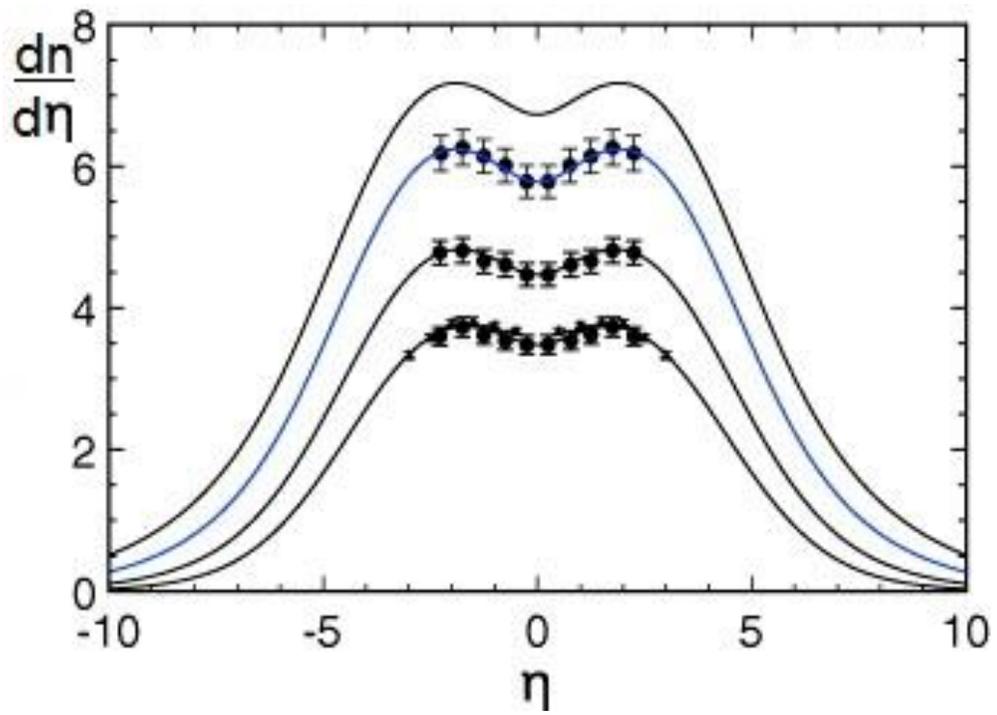

**Figure 2**

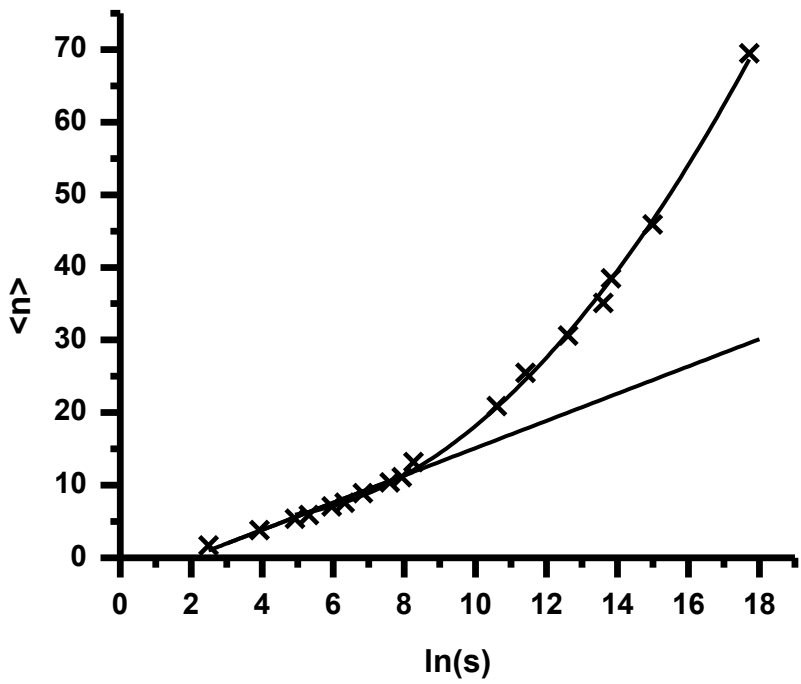

Figure 3

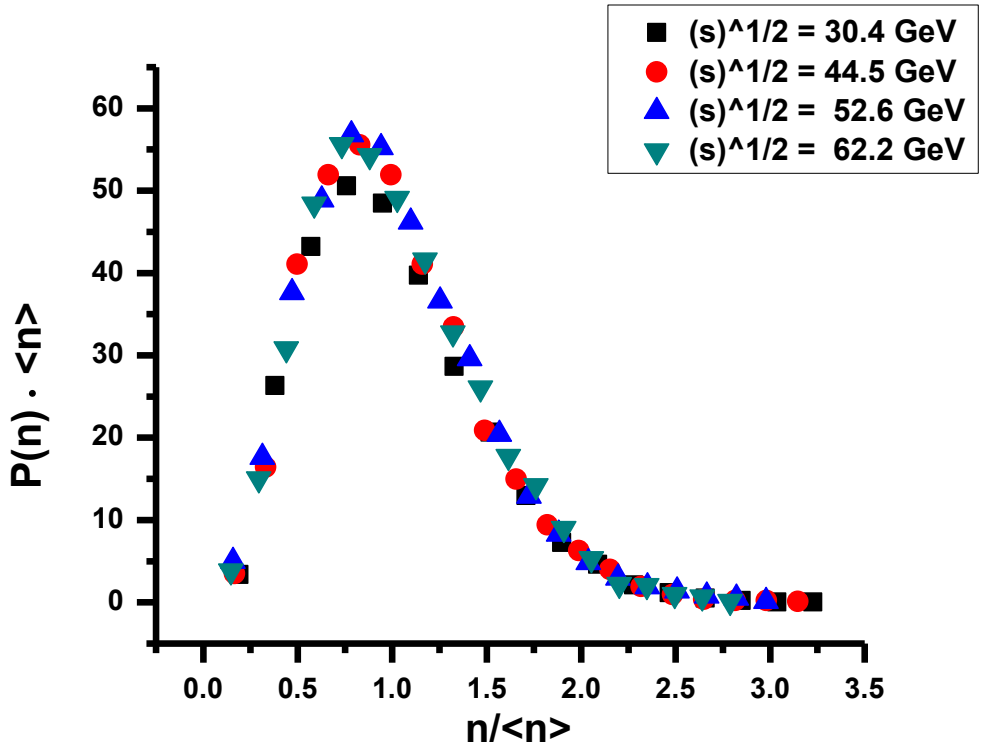

Figure 4

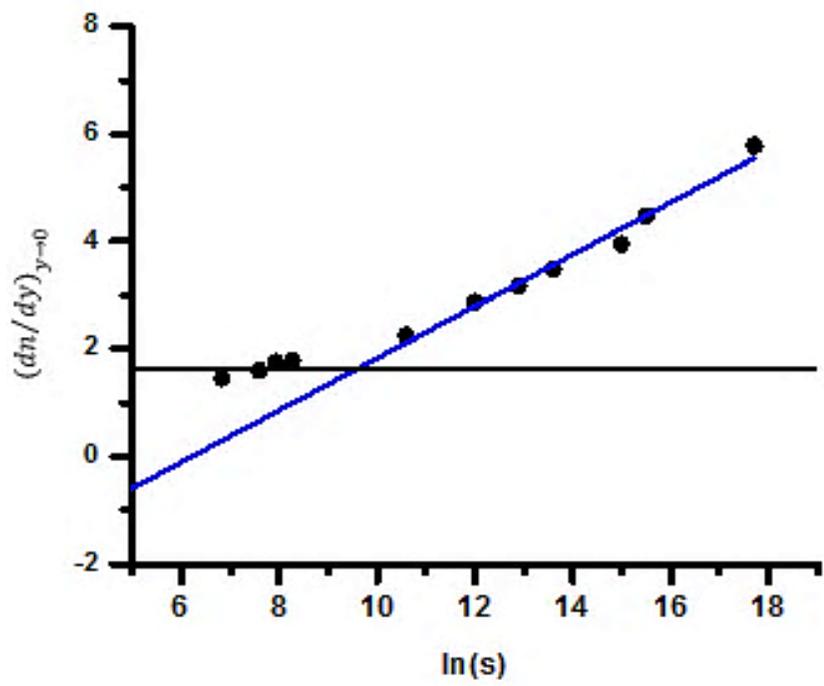

**Figure 5**

**Figure Captions**

Figure (1): the pseudo-rapidity distributions (dn/dη) of the created particles in p-p interactions at energies √s=23.6, 53, 200 and 410 GeV.

Figure (2): the pseudo-rapidity distributions (dn/dη) of the created particles in p-p interactions at energies √s=0.9, 2.36, 7 and 14 TeV. Blobs are experimental points while solid lines are calculations according to the RDM.

Figure (3): the mean multiplicity of created particles, <n>, as a function of ln (s). Crosses are experimental data, the straight line is the fit with the low energy part of the data and the curve is the fit with whole data.

Figure (4): the KNO scaling for charged particle multiplicity distributions at the ISR energies.

Figure (5): the dependence of the particle density per unit rapidity interval in central rapidity region, on √s of the interacting protons.